\documentclass[
    ,final            
  ]
  {aipproc}

\layoutstyle{6x9}

\usepackage{amssymb}
\usepackage{graphicx}

\begin{document}

\title{The Cosmological Slingshot Scenario: a Stringy Proposal for the Early Time Cosmology\footnote{Based on \cite{Germani:2006pf} and \cite{Germani:2007ub}.}}


\author{Cristiano Germani}{
  address={SISSA and INFN\ - \ via Beirut 4, 34014 Trieste, Italy\ - \ germani@sissa.it\\~}
}

\author{Nicol\'{a}s Grandi}{
  address={IFLP, CCT La Plata, CONICET - Dto. de F\'{\i}sica, UNLP \\ Casilla de Correos 67, CP1900 La Plata, Argentina\ - \ grandi@fisica.unlp.edu.ar\\~}
}

\author{Alex Kehagias}{
  address={Department of Physics, National Technical University of Athens \\ GR-15773, Zografou,
Athens, Greece
\ - \ kehagias@central.ntua.gr}
}

\begin{abstract}
In the Cosmological Slingshot Scenario, our Universe is a $D3$-brane that extends in the $4d$ noncompact directions of a warped
Calabi-Yau compactification of IIB Supergravity. Early time cosmology corresponds to a period in which the brane moves inside
a warped throat where a non-vanishing angular momentum ensures that the trajectory of the brane has a turning point.
The corresponding induced metric on the $D3$-brane experiences a cosmological evolution with a bounce. In this framework,
the homogeneity, flatness, and isotropy problems of standard cosmology might be avoided. The power spectrum of primordial
perturbations of the brane embedding can be found and it is shown to be in agreement to WMAP data.
\end{abstract}

\maketitle


\section{Introduction}

The Cosmological Slingshot Scenario \cite{Germani:2006pf,Germani:2007ub,Germani:2007uc}
is a proposal for the cosmic early-time evolution in
the String Theory context. According to that, our Universe is a
$D3$-brane moving in a String Theory background of the form
${\cal M}^4\times K^6$. ${\cal M}^4$ is a warped Minkowskian
space-time and $K^6$ is a compact Calabi-Yau (CY) space. The
latter includes a throat sourced by a stack of a large number
($N$) of other $D3$-branes. The Slingshot is characterized by a
non-trivial orbital motion of the Universe in the compact space
around the stack of $D3$-branes. If back-reaction can be neglected
(probe brane approximation), a brane observer measures a $4d$ metric
induced in terms of the brane embedding, that defines a cosmological brane
evolution commonly called Mirage Cosmology \cite{Kehagias:1999vr}.

The early-time evolution ({\it i.e.} well before nucleosynthesis)
corresponds to the motion of the $D3$-brane deep into the throat (Slingshot era) moving towards
the hat of the compact space. Since $N$ is taken to be large, close to the
stack the probe brane approximation can be used. The late-time cosmology starts
when the $D3$-brane reaches the hat of the CY, the probe brane approximation breaks down and local gravity
{\it \`a la} Randall-Sundrum \cite{Randall:1999vf,Shiromizu:1999wj} dominates the cosmological evolution.
Under this approximation, the
Slingshot brane observer experiences a non-singular bouncing
cosmology. Additionally, as we shall show later on, the
Standard Cosmology problems ({\it i.e.} homogeneity, isotropy and flatness) might be avoided in the brane induced cosmology.

\section{The Cosmological Slingshot Scenario}
\label{sKS}
\subsection{Setup}
To make the discussion concrete, we will consider a probe $D3$-brane moving in a throat of a Calabi-Yau (CY) compact manifold, whose metric and Ramond-Ramond $5$-form can be written as
\begin{eqnarray}
ds^2=h^{-\frac{1}{2}}ds^2_{\mbox{\tiny ${\cal M}_4$}} +
h^{\frac{1}{2}}\left(dr^2+r^2 ds^2_5)\right)\ ,\ \ \ \ \ \ \ C_{0123}=1-\frac1h\,,
\label{metrik}
\end{eqnarray}
where $h$ is a function of $r$ only, and $ds^2_5$ is the base manifold characterizing the remaining part of the transverse space.

The dynamics of a probe brane is governed by the Dirac-Born-Infeld
action with the Wess-Zumino coupling
\begin{eqnarray}S_{DBI}+S_{WZ}= -T_3\int \sqrt{-g_i}\,d^4\xi - T_3\int C_{(4)}
\, . \label{actionn}
\end{eqnarray}
We assume that
all other fields on the brane are switched off and matter is
created later. The sign of the Wess-Zumino term has been chosen to
represent a $D3$-brane and $T_3=1/{(2\pi)^3g_sl_s^4}$ is the tension of the probe.
The probe brane is extended
along the ${\cal M}_4$ directions, so that it looks like a point particle
moving in the transverse space. In the static gauge
the resulting induced
metric is
\begin{equation}
ds_i^2=h^{-1/2}\left[-\left(1-h(r'^2+r^2\Omega_5'^2){\phantom{z\!\!}}\right)dt^2+ d\vec x\cdot d\vec x\right]\ ,
\label{induced}
\end{equation}%
where a prime $(')$ denotes a derivative with respect to $t$ and we have assumed that the only non-vanishing transverse momenta are in the $r$, $\Omega_5$ directions, $\Omega_5$ representing an angle in the transverse space. Replacing this induced metric into the brane action (\ref{actionn})
we get
\begin{eqnarray}
S&=&
-T_3V_3
\int dt \,\left[\frac1h\sqrt{1-h\left(
r'^2+
r^2\Omega_5'^2
\right)}+1-\frac1h\right]\ , \label{dbikt}
\end{eqnarray}
where $V_{3}$ is the un-warped volume of the directions parallel to the probe. The resulting equations of motion have first integrals provided by
\begin{equation}
U = \frac1h\left[\frac1{ \sqrt{1-h(r'^2+r^2\Omega_5'^2)} }-1\right]
\,,\ \ \ \ \ \ \ \
J= \frac {r^2}{\sqrt{1-h(r'^2+r^2\Omega_5'^2)}}\,\Omega_5'\ ,
\label{enn}
\end{equation}
that can be inverted to get
\begin{equation}
r'^2=\frac1{(1+hU)^2}\left(2U-\frac{J^2}{r^2}+hU^2\right)\, .
\label{pin}
\end{equation}
The motion will take place at the values of $r$ that make this expression positive. Moreover wherever the expression in parenthesis vanishes, the trajectory will have a turning point.

To interpret the induced metric (\ref{induced}) as the cosmology experienced by an observer on the brane, we need to define the cosmic time according to
\begin{equation}
d\tau=h^{-1/4}\sqrt{1-h(r'^2+r^2\Omega_5'^2){\phantom{z\!\!}}}dt\,,
\end{equation}
in terms of which the metric (\ref{induced}) is identified as a Friedman-Robertson-Walker metric with scale factor
\begin{equation}
\label{scale}
a=h^{-1/4}(r) \ .
\end{equation}
When $r(t)$ has a turning point, it is easy to see that the same happens to $a(\tau)$, generating a
nonsingular bouncing cosmology. A Friedmann equation can be written for such cosmology by changing to cosmic time in Eq.(\ref{pin}) and dividing the result by $a^2$
\begin{equation}
H^2 =\frac1{16} \,a^6h_r^2(a)  \left[2U-\frac{J^2}{r^2(a)}+\frac{U^2}{a^4}\right] \ ,
\label{hub}
\end{equation}
where $H=\dot a/a$ is the (mirage) Hubble constant (here $\dot a=\partial_\tau a$).

The model is completed by  smoothly pasting this Mirage era to
a local gravity driven late evolution when the brane reaches the
top of the CY and gravity becomes localized {\em \'{a} la}
Randall-Sundrum \cite{Randall:1999vf,Shiromizu:1999wj}. There, the standard late time evolution of the observed Universe is supposed to be well reproduced by the brane dynamics.
This assumption involves a transition from a mirage dominated era with a moving brane without any matter,
into a local gravity dominated era with an static brane and matter fields excited on it. This transition
has to be understood as an analogous of the reheating process in standard inflationary models. It entails
a dynamical mechanism under which the kinetic energy of the brane is passed to matter fields. The
description of this dynamics as well as the robustness of our predictions for physical observables is an
open point that is left for future research.

\vspace{5pt}
A concrete example of the above proposed situation is given by a
probe moving in the $AdS_5\times S_5$ background. In it, the metric
takes the form  (\ref{metrik}) with
\begin{equation}
h_{AdS}=\frac{L^4}{r^4}\, , \ \ \ \ \ \ \ \ L^4=4\pi l_s^4 N g_s\, ,
\end{equation}
where  $l_s$ is the string length and $g_s$  is the
string coupling. The supergravity approximation is
valid as long as the curvature radius of the solution is large
compared to the string length $l_s$. String perturbation theory
on the other hand requires $g_s\ll 1$.

A close look to Eq.(\ref{pin}) shows that, in the present case, the second factor is a quadratic function of the variable $r^{-2}$, that will have a root as long as its discriminant is positive $J^4-8U^3L^4>0$. For values of $r$ larger that that of the root the function is positive. Then whenever this inequality is satisfied, the probe orbit has an inner turning point.

The resulting scale factor (\ref{scale}) reads
\begin{equation}
a_{AdS}=\frac rL\,.
\label{Ads}
\end{equation}
In this background, the induced Friedmann equation (\ref{hub}) becomes
\begin{equation}
H_{AdS}^2=\frac1{L^2}\left[\frac{2U}{a^4_{AdS}}-\frac{(J/L)^2}{a^6_{AdS}}+\frac{U^2}{a^8_{AdS}}\right]\ ,
\label{hub2}
\end{equation}%
where $H_{AdS}=\dot a_{AdS}/a_{AdS}$ is the (mirage) Hubble constant. Since when $J^4-8U^3L^4>0$ the orbit has a turning point, in that case the corresponding cosmology has a bounce.

\vspace{7pt}
Another example is that of a probe brane motion in a Klebanov-Strassler (KS) throat \cite{Klebanov:2000hb}. In the region far from the tip of the throat, KS geometry can be well approximated by the Klebanov-Tseytlin (KT) metric \cite{Klebanov:2000nc} that takes the form (\ref{enn}). The warp factor reads
\begin{equation}
h_{KS}=\frac{L^4}{r^4}\ln(\frac r{r_s})\, , \ \ \ \ \ \ \ \ L^2=(9/\sqrt8)\,
l_s^2 M g_s\, ,
\end{equation}%
where $r_s$ is proportional to the radius of the blown up sphere at the tip of the cone. To trust the KT approximation we need to ensure that the probe brane will never reach $r\simeq r_s$. This will be the case iff  $r_s$ lies inside the forbidden region $r'^2<0$. Going back to equation (\ref{pin}) and evaluating it at $r_s$, we see that this is true when $2r_sU-J^2<0$, the probe motion having a turning point at some value of $r$ larger than $r_s$.

The resulting scale factor is
\begin{equation}
a_{KS}=\frac{r}{L} \,\ln^{-1/4}(\frac r{r_s})\, ,
\label{ver}
\end{equation}
and, under the assumption $2r_sU-J^2<0$, it corresponds to a bouncing cosmology.

An important ingredient in our argument is that the scale
factor for a brane moving in a KS throat (\ref{ver}) can be rewritten as a conformal re-scaling of the corresponding
scale  factor for a brane moving in $AdS_5$. Indeed, when written in conformal time, the induced metric on the brane reads
$ds^2_i= a^{-2}ds^2_{\mbox{\tiny ${\cal M}_4$}} $
and we can write
\begin{equation}
a_{ KS} = \Omega(a_{AdS})\,a_{AdS}\,, \ \ \ \ \ \ \ \ \mbox{with}\ \Omega(a_{AdS}) = \log^{\!\!-1\!/4}\!(a_{AdS}L/r_s)
\label{ads}\ .
\end{equation}%
It should be kept in mind that our approximations are valid whenever $a_{AdS}\gg r_s/L$.
Under such a re-scaling, the Hubble constant changes as
\begin{equation}
H_{KS} = \left(1+a_{AdS}\frac{d\ln\Omega}{da_{AdS}}\right)H_{AdS}
\label{hubblee}\ .
\end{equation}%

\subsection{The Problems of Standard non-Inflationary Cosmology}
We are now ready to study how standard cosmological problems are solved in the Slingshot scenario.

\vskip.2cm
\label{intro}
{\noindent\bf{A. Homogeneity.}} As explained above, in both the $AdS$ case and the KS throat
the probe brane experiences a bounce in the String frame. This immediately ensures that
homogeneity problem is solved. To check this explicitly, we write the co-moving horizon
\begin{equation}\Delta\eta = \int_{\eta_i}^{\eta_0}d\eta\, ,
\end{equation}%
where $\eta$ is conformal time and $\eta_i$ is its smallest value. To solve homogeneity problem it is required that $\Delta\eta > H_0^{-1}$.
Since we have $\eta_i\to-\infty$ due to the absence of a cosmological singularity, this condition is trivially satisfied.\vspace{.1cm}

\vskip.2cm
{\noindent \bf{B. Isotropy.}} In the $AdS$ case, mirage matter contributes to Friedmann equation (\ref{hub2}) with a term
$\rho\sim a^{-8}_{AdS}$. This term dominates over the shear $\rho_{shear}\sim a^{-6}_{AdS}$ at early times,
avoiding the chaotic behavior \cite{Erickson:2003zm}.
To check whether this is true in the KS case, we should verify that the corresponding
mirage contribution dominates over the shear.
The form of this contribution can be read from (\ref{hubblee}), and we can write the quotient
\begin{equation}
\sqrt{\frac{\rho_{shear}}{\rho}}\propto\left(1+a_{AdS}\frac{d\ln\Omega}{da_{AdS}}\right)^{-1}\frac {a_{AdS}}{\Omega^3}\ .
\label{shear}
\end{equation}
The proportionality constant in (\ref{shear}) parameterizes the anisotropic perturbations in the pre-bounce era.
It is simple to check that (\ref{shear}) is an increasing function of $a_{AdS}$ in the region $a \gtrsim e^{((2\sqrt{2}-1)/4)} \,r_s/L\simeq1.57 \,r_s/L$.
As we assumed that the Slingshot brane never approaches the tip of the KS throat, this condition is automatically satisfied.
Therefore, ${\rho_{shear}}/{\rho}$ decreases very rapidly close to the bouncing point in the pre-bounce era, solving the isotropy problem.

\vskip.2cm
{\noindent \bf{C. Flatness.}} The curvature contribution to the Hubble
equation\footnote{The details of how to include a curvature term in the mirage Hubble equation can be found in \cite{Germani:2006pf}.} can be disregarded if the quantity $|\Omega_{\mbox{\tiny Total}}-1|\,=\,{1}/{a^2H^2}$ passes through a minimum where it satisfies the phenomenological constraint
\begin{equation}
|\Omega_{\mbox{\tiny Total}}-1|_{min}<10^{-8}
\, .\label{cons}
\end{equation}

For the $AdS$ case, the above quantity evaluated at its minimum reads
\begin{equation}
|\Omega_{\mbox{\tiny Total}}-1|_{min}=
\frac{(J^2+\sqrt{J^4-6L^4U^3})^3}{4L^2U^2(J^4-4L^4U^3+J^2\sqrt{J^4-6L^4U^3})}
\simeq
\left(\frac{J}{2LU}\right)^2\!\!+\!{\cal O}\!\left(\!\frac{L^4U^3}{J^4}\!\right)
<10^{-8}.
\end{equation}
This condition is not a fine tuning in parameter space, but just a restriction to a two dimensional region. In this sense flatness problem might be alleviated in the Slingshot scenario.

For the KS case we have, after conformal re-scaling
\begin{equation}
|\Omega_{\mbox{\tiny Total}}-1|=\frac {f^2}{a_{AdS}^2H_{AdS}^2}\ ,
 \ \ \ \ \ \ \ \
 \ \ \ \ \ \ \ \
 f=\frac{4\ln(a_{AdS} L/r_s)}{4\ln (a_{AdS} L/r_s)-1}\ .
\label{f1}
\end{equation}
The KT approximation is valid for $r_{min}\gg r_s$; to fix ideas we will use $r_{min}>10^2\ r_s$. In this region we have $f={\cal O}(1)$ and decreasing in $a_{AdS}$. Consequently,
the flatness problem in the KS space might, in good approximation,
be alleviated by the same choice of parameters used in the $AdS$ case.

\section{Primordial Perturbations}
\label{sPert}
\subsection{The Hollands-Wald Mechanism}

In inflationary scenarios the primordial perturbations are produced by quantum
fluctuations of the inflaton field and are codified into its two point
correlation function in the vacuum state. However, these fluctuations are
over-damped by the expansion of the Universe at super-horizon scales.
At these scales then, the quantum state becomes characterized by a large occupation number and
the system collapses into a classical state. This classical state
represents a random spectrum of perturbations whose variance is given by the quantum correlations evaluated at the
quantum-to-classical transition point \cite{Liddle:2000cg}.

Let us now turn our attention into the mechanism proposed by Hollands and Wald in \cite{Hollands:2002yb}.
A perturbation of wavelength $\lambda$ smaller than a typical quantum scale, say $l_c$,
is in its quantum vacuum. In an expanding background, the wavelength of a perturbation grows in time ($\lambda\propto a$) and whenever $\lambda\sim l_c$, or in other words, as soon as the perturbation becomes macroscopic, wavelengths
bigger than the horizon scale collapse into a classical random state.
In the proposal of \cite{Hollands:2002yb}, the relevant fluctuations are so continuously ``created'' at ``super-horizon'' scales.
Thus, a coherent spectrum of classical perturbations is produced with variance given by matching the classical correlations with the quantum correlations at the quantum-to-classical transition point.

It has been suggested \cite{Li:1996rp} that a space-time uncertainty relation $\Delta X\Delta T\gtrsim l_s^2$ should be realized in String Theory, $\Delta X$ and $\Delta T$ representing the uncertainties in measuring space and time distances. Since the smallest length that can be probed in String Theory is the $11$-dimensional Planck length $\Delta X>l_{P^{11}}\sim g_s^{\mbox{\tiny$1/3$}}l_s$, we obtain that the smallest measurable time is  $\Delta T\gtrsim g_s^{\mbox{\tiny$-1/3$}}l_s$. The period of a wave propagating in a $D$-brane is $2\pi\omega^{-1}\sim \lambda$ and cannot be smaller than that $\Delta T$, implying
\begin{equation}
\lambda>l_s g_s^{-1/3}\ .\label{second}
\end{equation}
We therefore have a minimal wavelength for a perturbation on the brane as in the Hollands-Wald mechanism.
This strongly suggests to use such mechanism to study the cosmological perturbations in the Slingshot
model\footnote{
To be precise, the Hollands-Wald mechnism can be used only for
perturbations with wavenumbers $k\geq k_{min}$, where
$k_{min}=a_{min}l_c^{-1}$.
Perturbations with $k<k_{min}$ never enter in the quantum
region. These perturbations are therefore normalized in
the past infinity and their associated spectrum is
generically blue [13].
However, as the size of these perturbations can be taken to
be much larger than the Hubble horizon today [3], they can be
safely excluded from current CMBR observations.}$^,$\footnote{In the original
proposal of \cite{Hollands:2002yb} the perturbation was produced by
the same radiation which sets the CMB. However, as pointed out by
\cite{Kofman:2002cj}, the perturbation coming out from the horizon
today, was necessarily born when the energy density of radiation was
much bigger than the Planck energy, which makes the mechanism
unreliable. In the Slingshot instead, perturbations are created by
brane fluctuations in a regime in which the supergravity
approximation is still valid, and no extra quantum effect takes
place. Moreover, super-horizon causality is not required, since in
the Slingshot perturbations are overdamped at sub-horizon scales
$k<J/r^2$.}.

Technically, the mechanism explained above introduces a vacuum state
in which the perturbation is destroyed (coming from the pre-bounce era) and then created again (after the bounce) at the time $\eta_*$ in which the proper wavelength of the corresponding quantum mode reaches the value
\begin{equation}a(\eta_*)/k\equiv a_*/k = l_c\, . \label{lc}
\end{equation}%

We start by perturbing the embedding of the probe brane by writing $r = r(\eta)+\delta r(\eta,\vec x)$ and
$\Omega_5 = \Omega_5(\eta)+\delta \Omega(\eta,\vec x)$, where
$r(\eta),\Omega_5(\eta)$ are the solutions of the equations of motion
obtained from action (\ref{actionn}), written
as functions of the conformal time $\eta$.
In the non-relativistic approximation $hU\ll1$ we have $\eta\equiv t$
and
Eqs.(\ref{enn}) are integrated to
\begin{equation}r(\eta)=\sqrt{2U\,\eta^2+\frac{J^2}{2 U}}\, ,
~~~~~\Omega_5(\eta)=\arctan\left(\frac{2U}{J}\,\eta\right)\, .\label{8}
\end{equation}%
Note that we have a turning point at $r_{min}=J/\sqrt{2U}$.

In what follows, we will use as our variable the Bardeen potential $\delta \Phi_k = \delta r_k/r$ \cite{Durrer:1993db},
in terms of which the action (\ref{actionn}) can be expanded to quadratic
order in $\delta$'s and their derivatives, getting (in Fourier space)
\begin{equation}
\!S\!=\!T_3\sum_k\!\int
d\eta \left( \phantom{\frac12}\!\!\! \! \! \! \! \! \frac{r^2}2\! \left(\! {\delta
\Phi'}_k ^2 + \delta {\Omega}'^2_k -k^2( \delta\Phi_k^2
+\delta\Omega_k^2) \right)+ J\,\delta {\Omega}_k'\delta \Phi_k
-J\,\delta {\Omega}_k\delta \Phi'_k \right) . \label{dos}
\end{equation}%

\subsection{Power Spectrum and Spectral Index}

The canonical quantization procedure applied to the action (\ref{dos}) provide the normalized operators
\begin{eqnarray}
 &&\delta\hat\Phi_k=u_1 \hat a_1+u_2 \hat a_2+c.c.\,, \ \ \ \ \ \ \ \ \ \ \delta\hat \Omega_k=v_1 \hat a_1+v_2\hat a_2+c.c.\ ,
\label{quant}
\end{eqnarray}
where $a_i,a_i^\dag$ are standard annihilation and creation operators, and
\begin{eqnarray}
 &&\!\!\!\!\!
 u_1=\sqrt{\frac{U}{k T_3}}\, \frac{\eta}{r^2}\,e^{-ik\eta}\, ,
 ~~~~~~~~~\ \ \ \ \ \ \ \ u_2=\sqrt{\frac{1}{U k T_3}}\, \frac{J}{2r^2}\,e^{-ik\eta}\,,\\
 &&\!\!\!\!\!
 v_1=u_2=\sqrt{\frac{1}{U k T_3}}\, \frac{J}{2r^2}\,e^{-ik\eta}\,,
 ~~\ \ v_2=-u_1=-\sqrt{\frac{U}{k T_3}}\, \frac{\eta}{r^2}\,e^{-ik\eta}\ .
\label{quantII}
\end{eqnarray}

We are interested in the correlation of the Bardeen potential $\delta \hat \Phi$ at the time of creation $\eta_*$. Using the above formulas to define $r_*=r(\eta_*)$, it is straightforward to check that
\begin{equation}
\langle\delta\hat\Phi_k\delta\hat\Phi_{k'}\rangle=\delta_{k,k'}\frac{1}{2kT_3r_*^2}\ .
\end{equation}%

We will consider the transition  point of the quantum to the classical description in the region in which
$k\ll J/r^2$. In this limit, we can discard the $k^2$ term in the action (\ref{dos}) and write its classical solutions as
\begin{eqnarray}\label{solutions}
\delta\Phi_k=\frac{C_k}{2J}+A_k\sin\left(2\theta+\phi_k\right)\ ,\ \ \ \ \ \
\delta\Omega=-\frac{D_k}{2J}+A_k\cos\left(2\theta+\phi_k\right)\ ,
\end{eqnarray}
where $\theta=\Omega_5(\eta)-\Omega_5(\eta_*)$
and $\phi_k,C_k,D_k,A_k$ are constants of integration, that can be written as
\begin{eqnarray}
C_k=r^2\delta\Omega_k'+2J\delta\Phi_k\ ,\ \ \ \ \ \ \ \,
D_k=r^2\delta\Phi_k'-2J\delta\Omega_k\, ,
\nonumber\\
A_k=\frac{r^2}{2J}\left[\delta\Phi_k'\cos\left(2\theta+\phi_k\right)-\delta\Omega_k'\sin\left(2\theta+\phi_k\right)\right]\ .
\end{eqnarray}%

We now consider  initial conditions arising from the matching of the classical to the quantum system at the
time $\eta=\eta_*$. Therefore $C_k,D_k,A_k$ will be taken as Gaussian stochastic variables with
correlations $\langle...\rangle_c$ matching the quantum correlators $\langle...\rangle$ at $\eta=\eta_*$.

Using the quantum solutions described above at the matching point $\eta=\eta_*$ after a lengthly but
straightforward calculation we have
\begin{eqnarray}
&&\langle C_{k} C_{k'}\rangle_{c}=\delta_{k,k'}\frac{k^2r_*^2+2U}{2kT}\ ,\ \ \ \ \ \ \
\langle A_{k} A_{k'}\rangle_{c}=\delta_{k,k'}\frac{k^2r_*^2+2U}{8J^2kT}\ ,
\nonumber\\&&
\langle A_{k} C_{k'}\rangle_{c}=\delta_{k,k'}\frac{\sin\phi}{4kTJr_*^2}\{2J^2-2Ur_*^2-k^2r_*^4
-4JU\eta_*\cot\phi_k\}\ .
\end{eqnarray}
The  matching of
\begin{equation}\langle \delta\Phi_k\delta\Phi_{k'}\rangle_{c}=\langle \delta\hat\Phi_k\delta\hat\Phi_{k'}\rangle\ ,
\end{equation}%
requires $\phi_k=\pi/2$; this is the selection of positive frequencies.

In general correlators depend on time through $\theta$.
However in the region $k\ll J/r^2$, the oscillation rapidly stabilizes
in time when $2U\eta_{\mbox{\tiny asymp.}}>2\pi J$. We will consider this to happen well before the nucleosynthesis. At this time
then
\begin{equation}\delta\Phi_k=\frac{C_k}{2J}-A_k\cos(2\Omega_5(\eta_*))=\frac{C_k}{2J}+A_k\left(1-\frac{2r_{min}^2}{r_*^2}\right)\ .
\end{equation}%
Using the initial conditions found above we then get in the limit $k\ll J/r_*^2<J/r_{min}^2$
\begin{equation}\langle \delta\Phi_k\delta\Phi_{k'}\rangle_{c}\Big|_{\eta>\eta_{\mbox{\tiny asymp.}}}\simeq \frac{\delta_{k,k'}}{2kTr_*^2}\left[1-
\left(\frac{r_{min}}{r_*}\right)^2\right]\ ,
\end{equation}%
so the power spectrum of temperature fluctuations is
\begin{equation}P(k)\simeq\frac{1}{2k\,T_3\,r_*^2}\left[1-
\left(\frac{r_{min}}{r_*}\right)^2\right]\ . \label{power}
\end{equation}%
A consistency condition for the production of the perturbation is that $r_{min}<r_*$. So we see that in the limit
$r_{min}\ll r_*$ we obtain the power spectrum introduced in \cite{Germani:2006pf}.
\vskip.2cm
Since we assumed that a perturbation is created when its physical wavelength reaches a fixed value $l_c$, we have from Eq.(\ref{lc}), $k l_c= a_*$. In the $AdS$ metric Eq.(\ref{Ads}) implies $kl_cL=r_*$, resulting in the power spectrum
\begin{equation}
P(k)\simeq\frac{1}{2\,T_3(l_cL)^2 \,k^3}\left[1-\frac{r_{min}^2}{(l_c
    L)^2\,k^2}
\right]\, , \label{powerAds}
\end{equation}%
for which the scalar spectral index $n_s-1=d\ln (k^3P(k))/d\ln k$ reads
\begin{equation}
n_s\simeq1+\frac{2}{\frac{(l_cL)^2}{r_{min}^2}k^2-1}\,  \label{indexAds}
\end{equation}%
and we see that the flat spectrum found in \cite{Germani:2006pf} is blue-shifted by the subsequent time evolution.
\vskip.2cm
In the Klebanov-Tseytlin (KT) metric on the other hand, the condition
$kl_c=a_*$ is solved by
$r_* = r_s e^{{-\,W_{-1}(-\zeta)}/4}$ where $\zeta=4(r_s/Ll_ck)^4\leq
e^{-1}$
and $W_{-1}(x)$ is the negative branch of Lambert's $W$-function.
Then the power spectrum (\ref{power}) is explicitly written as
\begin{equation}P(k)=\frac{1}{2T_3\, k \, r_s^2} \,e^{{\frac12
W_{-1}(-\zeta)}}\!\left(1\!-\!\left(\frac{r_{min}}{r_s}\right)^2e^{\frac12
    \,
W_{-1}(-\zeta)}\right)\,,
\end{equation}%
whereas the scalar spectral index turns out to be
\begin{eqnarray}
n_s &=& 1 + \frac{2}{1+W_{-1}(-\zeta)}\!\left(1-\frac{W_{-1}(-\zeta)}{1\!-\!({r_s}/{r_{min}})^2e^{-\frac12W_{-1}(-\zeta)}}\right)
\nonumber\\
&\simeq& 1 +
\frac{2}{\ln(\zeta)}-\frac{2\sqrt\zeta}{\sqrt\zeta-({r_s}/{r_{min}})^2}\,
,
\label{hi}
\end{eqnarray}%
where
the expansion of the Lambert W function for small argument
$W(-\zeta)\simeq \ln(\zeta)+\cdots$ was used in the second line.
Since in this limit $\ln(\zeta)<0$, the first correction on $n_s$ is negative.
On the other hand, the second correction is red or blue according to
the sign of its denominator. It will be negative whenever
\begin{equation}
\sqrt\zeta>{r_s^2}/{r_{min}^2}\, ,
\label{hola}
\end{equation}%
from which we immediately see that long wavelengths are red-shifted.

If the last term is instead positive, then $\sqrt\zeta<{r_s^2}/{r_{min}^2}$ and the overall sign of the correction has to be evaluated
taking into account the joint contribution of both terms in (\ref{hi}). After some manipulations we find that the correction is red whenever
\begin{equation}{\sqrt\zeta}\left(1-2\,{\log\!\sqrt\zeta}\right)<({r_s}/{r_{min}})^2\, ,
\end{equation}%
from which we conclude that short wavelengths are also red-shifted, and there is an intermediate range of wavelengths that is blue-shifted.

\section{Back-Reaction and Effective 4d Theory}

The 4D effective theory for warped compactifications of IIB supergravity with (static) D-branes
has been derived by a perturbative approach in \cite{GM}, and
by a gradient expansion method in \cite{KK2}. Using these results, in \cite{Germani:2007ub} the following effective 4d
Lagrangian describing Slingshot cosmology has been found
\begin{equation}
\! \!\! \! \! \! \!\! \! \! \! \!\!\!\!\!\!\!\!\! S_{brane}=\frac{L^2}2\int d^4x \sqrt{-g}\left[\left(\frac{1}{\kappa^2r^2}-
\frac{T_3}{6 N}\right)R+\frac{6}{\kappa^2}\frac{(\nabla r)^2}{r^4}-\frac{T_3}{N}(\nabla\Omega_5)^2\right]\ .
\label{eff2}
\end{equation}
The resulting equations of motion, specialized to a Friedmann-Robertson-Walker background with scale parameter $a(\eta)$, result into the following set of equations
\begin{eqnarray}
&&\frac{r''}{r}+2\frac{r'}{r}\left(\frac{a'}{a}-\frac{r'}{r}\right)+{\Omega'^2_5}=0\ ,
\ \ \ \ \ \ \frac{d~}{d\eta}\!\!\left(a^2\Omega'_5\right)=0\,,
\nonumber\\
&&\frac{T_3\kappa^2}{N}\left(\frac{a'^2}{a^2}+\Omega_5'^2\right)=
\,\frac{6}{r^2}\!\left(\frac{a'}{a}-\frac{r'}{r}\right)^2\!.
\end{eqnarray}
As can be checked by direct substitution, an exact solution of the full system is
\begin{eqnarray}
a=\frac1L\sqrt{\frac{J^2}{2U}+{2U}\eta^2}\ ,\ \ \ \ \ \ \Omega_5'=\frac{J}{L^2}\,\frac1{a^2}\,, \ \ \ \ \
r=\frac{a\,L}{1+\kappa\sqrt{\frac{UT_3}{3N}}\,\eta}\ .
\end{eqnarray}
We obtained the same scale factor evolution as in the mirage
approximation in
\cite{Germani:2007ub}, but now considering local gravity
back-reactions. The only
difference from the mirage approximation is on the $r$ evolution, due
to the
denominator $1+ \kappa\sqrt{UT_3/3N}\eta$. Since
$\kappa\sqrt{UT_3/3N}$
is supposed to be small,
the mirage approximation breaks down at very late or very early times,
{\it i.e.}
when the brane leaves the throat, as expected.
When this denominator is taken into account, the system is no longer time symmetric and there is a field singularity at the time in which the denominator vanishes. However, since the extra-dimensional space is compact $r<r_{max}$, where
$r_{max}$ defines the cut-off of the compact space, this singularity is just fictitious, and the effective action cannot be trusted when $1+ \kappa\sqrt{UT_3/3N}\eta\simeq 0$. There a description {\it \'{a} la} Randall-Sundrum \cite{Shiromizu:1999wj} must be used.

\section{Summary}

The Cosmological Slingshot scenario provides an example of a bouncing
cosmology in which the dynamics of the bounce is under control. It
provides alternative solutions to the problems of standard cosmology,
and the resulting perturbations spectrum is in agreement with WMAP
data. In the way it has been presented here it still has
to be checked that all the constraints, that appeared during our
calculations due
to our approximations and to phenomenological inputs, are mutually
compatible.
Even if in \cite{Germani:2006pf} some of these cross checks have been performed successfully, a complete cross-checking is still needed, and it may be the case that all the problems of standard cosmology cannot be solved at the same time. For example, to make our solution of flatness problem compatible with perturbation spectrum, a very strong lower bound on the conserved quantity $U$ is needed. The model is currently under research.

\section*{Acknowledgements}

C.G. wishes to thank Cliff Burgess and Toni Riotto for useful discussions. This work is partially supported by the European Research Training Network MRTN-CT-2004-005104 and the PEVE-NTUA-2007/10079 programme. N.E.G. wants to thank Department of Particle Physics of Santiago de Compostela University for hospitality during part of this work.

\end{document}